\documentclass[12pt]{article}
\usepackage{epsfig}
\usepackage{amsmath}
\usepackage{hhline}
\usepackage{amssymb}
\usepackage{times}

\newlength{\dinwidth}
\newlength{\dinmargin}
\def\gsim{\,\lower.25ex\hbox{$\scriptstyle\sim$}\kern-1.30ex%
\raise 0.55ex\hbox{$\scriptstyle >$}\,}
\def\lsim{\,\lower.25ex\hbox{$\scriptstyle\sim$}\kern-1.30ex%
\raise 0.55ex\hbox{$\scriptstyle <$}\,}

\begin{document}

\pagestyle{empty}
\begin{titlepage}

\noindent
\begin{flushleft}
{\tt DESY 03-082    \hfill    ISSN 0418-9833} \\
{\tt July 2003}                  \\
\end{flushleft}

\vspace*{3cm}

\begin{center}
  \Large
{\bf Multi--Electron Production at High Transverse Momenta \\
in {\boldmath $ep$} Collisions at HERA }

  \vspace*{1cm}
    {\Large H1 Collaboration} 
\end{center}

\begin{abstract}

\noindent
Multi--electron production is studied at high electron transverse momentum 
in positron--~and~electron--proton collisions using the H1 detector at HERA. The data correspond to an integrated luminosity of 115 pb$^{-1}$. Di--electron and tri--electron event yields are measured. Cross sections are derived in a restricted phase space region dominated by photon--photon collisions. In general good agreement is found with the Standard Model predictions.
However, for electron pair invariant masses above 100~GeV,
three di--electron events and three tri--electron events are observed, compared to   Standard Model expectations of $0.30\pm0.04$ and $0.23\pm0.04$, respectively.
\end{abstract}

\vfill
\begin{center}
{\it To be submitted to Eur. Phys. J.}
\end{center}

\end{titlepage}
\begin{flushleft}

A.~Aktas$^{10}$,               
V.~Andreev$^{24}$,             
T.~Anthonis$^{4}$,             
A.~Asmone$^{31}$,              
A.~Babaev$^{23}$,              
S.~Backovic$^{35}$,            
J.~B\"ahr$^{35}$,              
P.~Baranov$^{24}$,             
E.~Barrelet$^{28}$,            
W.~Bartel$^{10}$
% ...

\bigskip{\it
 $ ^{1}$ I. Physikalisches Institut der RWTH, Aachen, Germany$^{ a}$ \\
% ...
$ ^{4}$ Inter-University Institute for High Energies ULB-VUB, Brussels;
          Universiteit Antwerpen (UIA), Antwerpen; Belgium$^{ c}$ \\
% ...
$ ^{10}$ DESY, Hamburg, Germany \\
% ...
$ ^{23}$ Institute for Theoretical and Experimental Physics,
          Moscow, Russia$^{ l}$ \\
 $ ^{24}$ Lebedev Physical Institute, Moscow, Russia$^{ e}$ \\
% ...
$ ^{28}$ LPNHE, Universit\'{e}s Paris VI and VII, IN2P3-CNRS,
          Paris, France \\
% ...
$ ^{31}$ Dipartimento di Fisica Universit\`a di Roma Tre
          and INFN Roma~3, Roma, Italy \\
% ...
$ ^{35}$ DESY, Zeuthen, Germany \\
% ...
$ ^a$ Supported by the Bundesministerium f\"ur Bildung und Forschung, FRG,
      under contract numbers 05 H1 1GUA /1, 05 H1 1PAA /1, 05 H1 1PAB /9,
      05 H1 1PEA /6, 05 H1 1VHA /7 and 05 H1 1VHB /5 \\
% ...
$ ^c$ Supported by FNRS-FWO-Vlaanderen, IISN-IIKW and IWT \\
% ...
$ ^e$ Supported by the Deutsche Forschungsgemeinschaft \\
% ...
$ ^l$ Partially Supported by Russian Foundation
      for Basic Research, grant    no. 00-15-96584 \\
\smallskip
% ...
}
\end{flushleft}
\newpage

\pagestyle{plain}
\section{Introduction}
In this paper we describe  the first measurement 
of multi--electron production  at high transverse momentum ($P_T$) 
in electron\footnote{In this paper the term
``electron'' is used generically to refer to both electrons and positrons.}--proton interactions
at HERA.  Within the Standard Model (SM), 
the production of multi--lepton events in $ep$ collisions proceeds
mainly through photon--photon interactions; 
photons radiated from the incident electron and proton interact to produce
a pair of leptons,
$\gamma\gamma\rightarrow \ell^+ \ell^-$~\cite{verm}. 
At large invariant masses, multi--lepton production may be sensitive to new phenomena, 
 for instance the production of a doubly charged Higgs boson~\cite{newph1} or processes involving bileptons, generic bosons carrying two units of lepton number~\cite{newph2}.
\par
The analysis presented here is based on  data recorded in 1994--2000 by the H1 experiment. Electrons of 27.6~GeV collided with protons of 820 or 920~GeV, corresponding to  centre--of--mass energies $\sqrt{s}$ of 301~GeV or 319~GeV, respectively.
The total integrated luminosity of 115.2~pb$^{-1}$ consists of 
36.5~pb$^{-1}$ of $e^+p$ collisions recorded at 
$\sqrt{s}$ = 301~GeV and 65.1~pb$^{-1}$ recorded at 319~GeV, as well as  
13.6~pb$^{-1}$ of $e^-p$ collisions recorded at $\sqrt{s}$ = 319~GeV. 
A related study of muon pair production is presented in~\cite{multim}.

\section{Standard Model Processes and their Simulation}
\label{theory}
\begin{figure}[p] 
  \begin{center} \vspace*{-.5cm}
      \epsfig{file=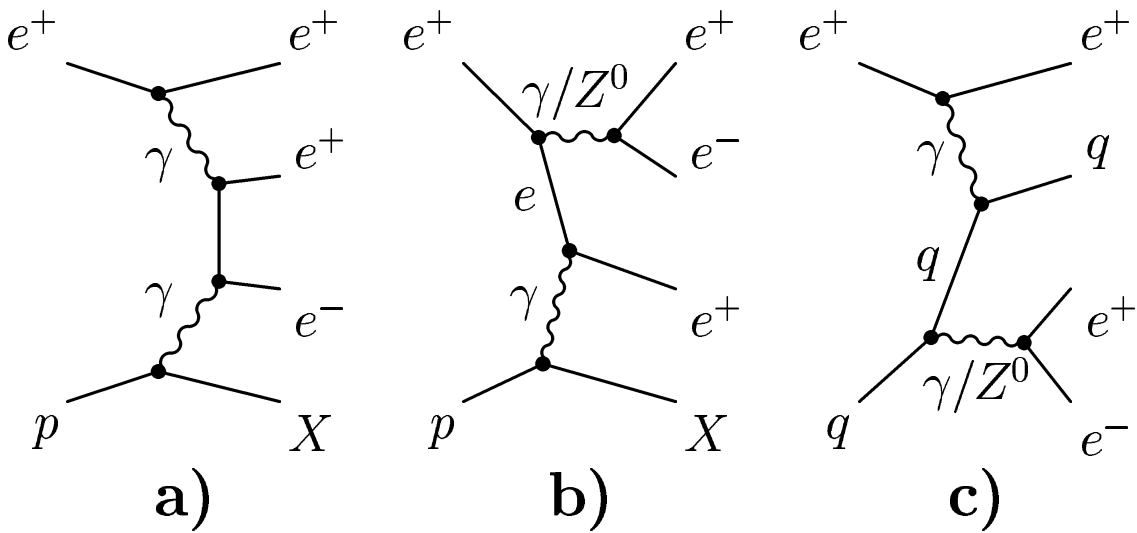,width=15cm}
  \end{center} \vspace*{-.3cm}
  \caption{
 The main processes involved in lepton pair production. Examples of Feynman diagrams  are shown for:  a) photon--photon interaction;  b) and c) $\gamma/Z^0$ boson conversion. 
The hadronic final state (X) can be  a proton  (elastic process)  or a higher mass system (quasi-elastic and inelastic processes).
}
  \label{diagrams}
\end{figure}

The main SM processes contributing to multi--electron production at HERA
are summarised in figure~\ref{diagrams}. 
The dominant contribution, shown in diagram~\ref{diagrams}a, is due to electron pair production through
the interaction of two photons radiated from the incident electron and proton.
Electron pairs can also originate from internal conversion of a photon ($\gamma$) or a $Z^0$ boson, radiated  either from the electron line (diagram~\ref{diagrams}b) or from the quark line (diagram~\ref{diagrams}c).   
The pole due to the electron propagator in diagrams~\ref{diagrams}a~and~\ref{diagrams}b 
corresponds to an $e^+e^-$ interaction in which one of the electrons is emitted from a photon radiated 
from the proton.  This mechanism is called the Cabibbo--Parisi process. Its contribution is one order of magnitude lower than the photon-photon contribution, except at high transverse
momentum, where it is more significant due to  $s$-channel $Z^0$ boson production (diagram~\ref{diagrams}b). 
In diagram~\ref{diagrams}c, the  pole  due to the quark propagator corresponds to the Drell--Yan process, $q\bar{q}\rightarrow e^+e^-$.
Its contribution is small compared with the photon--photon and Cabibbo--Parisi processes~\cite{romero}. 
\par
If the photon coupled to the incoming electron has a high virtuality, the incident electron can scatter through a large angle and with high transverse momentum. If the scattered electron is observed in  the detector, it is indistinguishable from the pair--produced electron of the same charge.
\par
The Monte Carlo generator GRAPE~\cite{grape} simulates lepton pair production in $ep$ collisions using  the full set of electroweak matrix elements at the tree level, with the exception of the Drell-Yan pole contribution.
GRAPE is based on the automatic Feynman graphs calculation program GRACE~\cite{GRACE}. 
Initial and final state radiation processes 
(QED and QCD parton showers) are simulated in the leading log approximation. 
The production of $\tau$ lepton pairs and their  subsequent electronic  decay
is also simulated with GRAPE and composes about 2\% of the multi--electron event sample.
 The Drell--Yan contribution is  simulated using the PYTHIA~\cite{PYTHIA} event generator. Its contribution is found to be very small in the  phase space relevant here and is neglected in the following. 
\par
In GRAPE the proton interaction  
is divided into in three phase space regions: elastic, quasi--elastic and inelastic.
In the case of elastic scattering, $ep\rightarrow ee^+e^-p$, the proton vertex is described in terms of dipole form factors. 
The quasi--elastic domain is defined by requiring that the mass of the hadronic final state $M_X<5$~GeV or that the virtuality of the photon coupled to the proton  $Q^2_p<1$~GeV$^2$.  
In the region $M_X<2$~GeV a resonance parameterisation~\cite{Brasse:1976bf} is used for the proton vertex. In the remaining quasi--elastic phase space, a fit to photoproduction and deep inelastic scattering data is used~\cite{Abramowicz:1997ms}. 
 The inelastic regime corresponds to electron--quark interactions with  $M_X>5$~GeV and $Q_p^2>1$~GeV$^2$.
In this case, the proton structure is parameterised using the  CTEQ5L parton distributions~\cite{Lai:1999wy}. The fragmentation and hadronisation processes are simulated using the SOPHIA program~\cite{Mucke:1999yb} in the quasi-elastic  and PYTHIA~\cite{PYTHIA} in the inelastic regime.
\par
The uncertainty attributed to the GRAPE calculation in this analysis is 3\%. This value results mainly from the~~uncertainties~~in~~the~~QED matrix~element~~calculation~~(1\%),~the parameterisation of the structure functions and the phase space separation between quasi-elastic and inelastic processes.
\par
The GRAPE prediction is cross--checked using the LPAIR generator~\cite{Baranov:yq}, 
which contains only the photon--photon process. When restricted to this process, the total and differential cross sections produced with LPAIR and GRAPE agree at the percent level. The additional diagrams in GRAPE
increase the predicted cross section by 10 \% on average in the phase space considered here. The increase is more pronounced (up to
30 \%) for di--lepton masses which are either very low (photon internal conversions) 
or around 90~GeV ($Z^0$ resonance production). 
\par 
The main experimental backgrounds to multi--electron production are 
processes in which, in addition to a true electron, one or more fake 
electrons are reconstructed in the final state. 
The dominant contribution arises from
neutral current Deep Inelastic Scattering (DIS) events ($ep\rightarrow eX$) in which, in addition to the scattered electron, hadrons or radiated photons are incorrectly identified as electrons. QED Compton scattering ($ep\rightarrow e\gamma X$) can also 
contribute if the  photon is misidentified as an electron. 
The DIS and elastic Compton processes are simulated using  the 
DJANGO~\cite{DJANGO} and WABGEN~\cite{Berger:kp} generators, respectively.
\par
All generated events are passed through the full GEANT~\cite{Brun:1987ma} based simulation of the H1 apparatus and are reconstructed using the same program chain as for the data. 
\section {Experimental Conditions} 
 
A detailed description of the H1 detector can be found in \cite{H1detector}.
The components essential for this analysis are described briefly
here.
\par
A tracking system consisting of central and forward\footnote{The origin of the H1 coordinate system is the nominal
$ep$ interaction point. The direction of the proton beam defines
the positive $z$--axis (forward direction). Transverse momenta are measured in the $x$--$y$ plane. Polar~($\theta$) and~azimuthal~($\phi$) angles are measured with respect to this reference system.
The pseudorapidity is defined as $\eta= -\log{\tan (\theta/2)}$.}
drift chambers is used to measure
charged particle trajectories and to determine the interaction vertex.  
The central tracker is composed of two concentric cylindrical drift chambers with an active detection region starting  at a radius of 22~cm.
The  angular range $37^\circ < \theta < 144^\circ$ is  covered by both  chambers. 
The inner drift chamber provides full acceptance for particles in the range $22^\circ < \theta < 159^\circ$. Particles at $\theta=20^\circ$ cross 83\% of its acceptance region.
Transverse momenta ($P_T$) are determined from the curvature of
the particle trajectories in a magnetic field of 1.15 Tesla.
The  central tracking system provides  transverse momentum measurements with a 
resolution  of $\sigma_{P_T}/P_T^2=5\times10^{-3}$ GeV$^{-1}$.
The tracking is complemented in the region $7^\circ < \theta < 25^\circ$ by a system of drift chambers perpendicular to the beam axis.
\par
Hadronic and electromagnetic final state particles
are absorbed in a highly segmented liquid argon calorimeter
\cite{h1cal} covering the range 
$4^\circ < \theta < 153^\circ$. The calorimeter is 5 to 8 hadronic interaction
lengths deep, depending on the polar angle and has an electromagnetic section which is 20 to 30 
radiation lengths deep.
Electromagnetic shower energies are measured with a precision of
$\sigma (E)/E = 12 \% / \sqrt{E/\mathrm{GeV}} \oplus 1\%$ and
 hadronic shower energies with a precision of
$\sigma (E)/E = 50 \% / \sqrt{E/\mathrm{GeV}} \oplus 2 \%$, as measured in test beams~\cite{h1calotestbeams}.
The electromagnetic energy scale is known to 0.7\% in the central region and  to 3\% in the forward region.
The hadronic energy scale is known to 2\%.
\par
In the backward region, energy measurements are provided 
by a lead/scintillating--fibre calorimeter\footnote{This device was installed in 1995, replacing a lead--scintillator
``sandwich'' calorimeter~\cite{H1detector}.}~\cite{SPACAL} covering
the range $155^\circ < \theta < 178^\circ$. 
The calorimeter system is surrounded by a superconducting coil
with an iron yoke which is instrumented with streamer
tubes. 
The electron and photon
taggers located downstream of the interaction point in the electron 
beam direction are used to determine the luminosity through the measurement of the  Bethe-Heitler  $ep\rightarrow e\gamma p$ process.
\par
The trigger used relies on  the liquid argon calorimeter signals and has an  efficiency  which is greater than 95\% for events in which an electron of energy above 10~GeV is detected.

\section{Data Analysis}
\subsection{Multi--electron event selection}
The multi--electron event selection is based on a
procedure which is designed to minimise the contribution of fake electrons,
while keeping a high efficiency for identifying true electrons and allowing reliable monitoring  of the overall selection efficiency.
\par 
As a first step, electron candidates with energies above 5~GeV are identified in the 
liquid argon and backward calorimeters, in the range $5^\circ < \theta < 175^\circ$. Electromagnetic showers 
are identified  with an efficiency of better than 98\%  using pattern recognition algorithms based on the geometric profiles expected for electrons. 
The remaining calorimeter clusters are attributed to hadronic activity and are combined into  jets using an inclusive
$k_T$ algorithm~\cite{ktalgo}, with a minimum jet transverse momentum of 4 GeV. 
Electron candidates are required to be isolated by demanding that they are separated from other electrons or jets by at least 0.5 units in the $\eta-\phi$ plane. In addition, the total hadronic energy within 0.75 units in $\eta-\phi$ of the electron direction is required to be below 2.5\% of the electron energy.

\par
In the region of angular overlap between the liquid argon calorimeter and the
central drift chambers ($20^\circ < \theta < 150^\circ$), the calorimetric
electron identification is complemented by tracking conditions. 
In this region it is required that a  high quality track be  geometrically matched to the electromagnetic cluster with a distance of closest approach to the cluster centre of gravity of less than 12~cm.
The starting radius of the measured track, defined as the distance between the first measured point in the central drift chambers  and the beam axis,
is required to be below 30~cm in order to reject photons that convert in the central tracker material beyond this radius. The transverse momentum of the associated track $P_T^{e_{tk}}$ and the calorimetric transverse momentum $P_T^{e}$ are required to satisfy the condition  $1/P_T^{e_{tk}}-1/P_T^e<0.02$~GeV$^{-1}$.  No other high quality track is allowed within 0.5 units in $\eta-\phi$ of the electron direction.
These additional constraints strongly
reduce the contribution of fake electrons from misidentified photons and hadrons. The resulting electron finding efficiency is 88\%. Electrons selected in this polar angular range are called ``central electrons'' hereinafter. 
\par
Due to the  higher material density in the forward region ($5^\circ < \theta < 20^\circ$) the electrons are more likely to shower and therefore no track conditions are required. 
The same applies in the backward region  ($150^\circ < \theta < 175^\circ$). 
 The forward electron energy threshold is raised to 10~GeV in 
order to reduce the number of fake electrons arising from hadrons in DIS  events.
\par
The final multi--electron event selection requires that there be two central electron candidates,
 of which one must have $P_T^e > 10$ GeV and the other
$P_T^e > 5$ GeV. Additional electron candidates are identified 
in the central and backward regions with $E^e>5$~GeV and in the forward region with $E^e>10$~GeV.  
The electron candidates are ordered
according to decreasing $P_T$, $P_T^{e_i} > P_T^{e_{i+1}}$.
\par
The selected events are classified as ``2e'' if
only the two central electron candidates are identified and ``3e'' if exactly one additional electron candidate 
is identified. 
A subsample of the ``2e'' sample, labelled ``$\gamma\gamma$'',
is selected in order to measure the pair production  
cross section in a well defined phase space region dominated by photon--photon collisions with low background.
In this subsample, the two electrons must be of opposite charge
and a significant deficit compared to the initial state 
must be observed in the difference $E - P_z$ of the energy and longitudinal momentum of all visible particles ($E-P_z<45$~GeV) 
\footnote{For fully contained events or events where only longitudinal momentum
along the proton direction ($+z$) is undetected, one expects 
$E - P_z = 2E_e^0 = 55.2$~GeV, where $E_e^0$ is the energy of the incident
electron. If the scattered electron is undetected, the threshold $E-P_z<45$~GeV corresponds to a cut on the fractional energy loss $y=(E-P_z)/2E_e^0<0.82$.}.
These two conditions ensure that the incident electron
is lost in the beam pipe after radiating a quasi--real photon of
squared four--momentum $Q^2$ lower than 1 GeV$^2$.

\subsection{Background studies}
\label{bg}
\par
\begin{figure}[htb]
  \begin{center}
     \epsfig{file=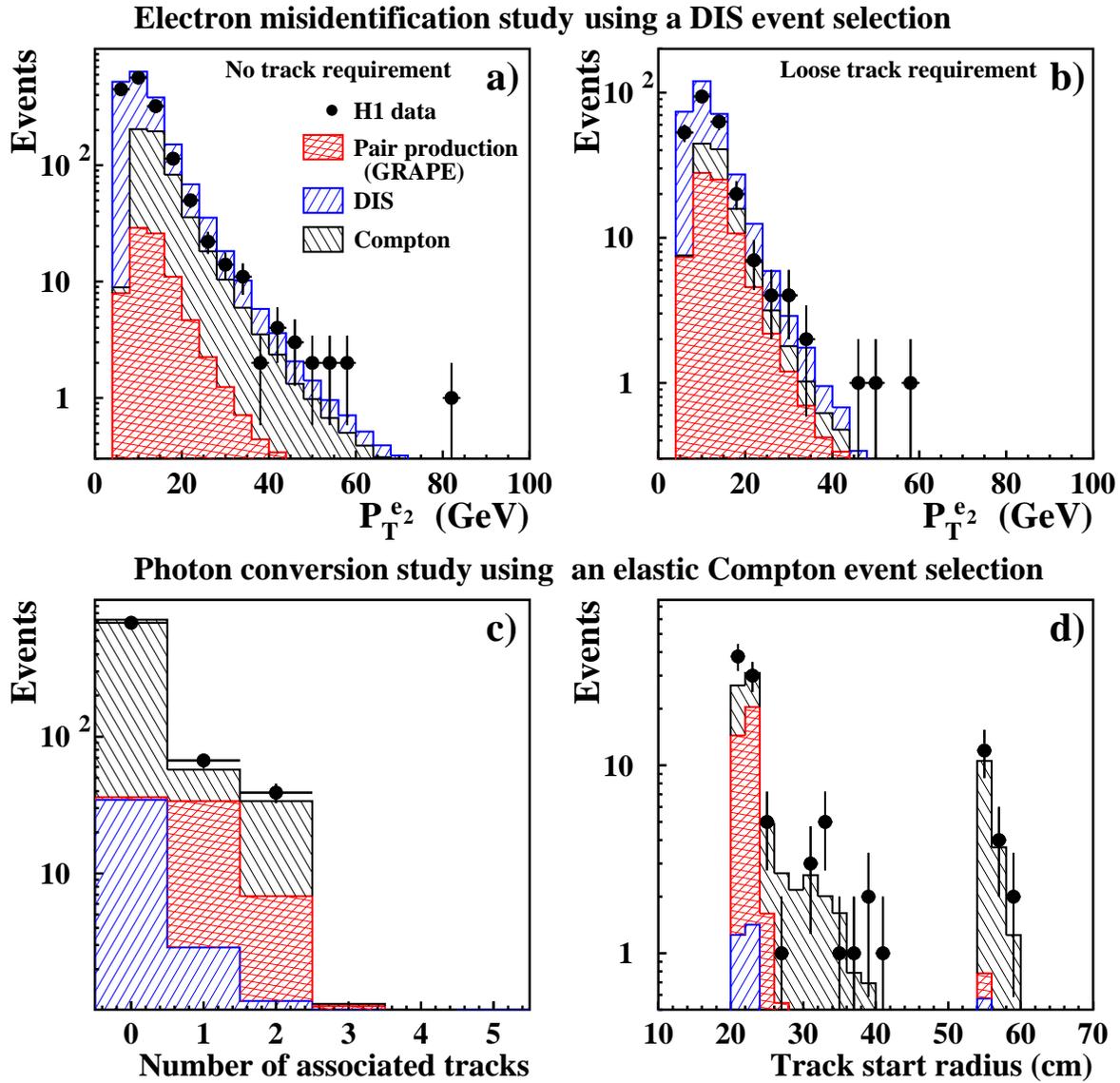,width=16cm}%
  \end{center}
  \caption{ 
(a and b) Distributions associated with misidentified electrons for a DIS event selection.
Transverse momentum spectrum of second electrons identified 
with either no track requirement (a) or a loose track requirement (b)
compared with expectations. (c and d)
Study of photon conversions using an elastic Compton event selection:
number (c) and starting radius (d) of the tracks associated 
with the photon candidates compared with  expectations. 
}
  \label{control}
\end{figure}

DIS and Compton processes can contribute to the selected multi--electron sample if a photon or a particle from the hadronic final state is misidentified as an electron. In order to quantify the uncertainty on the background prediction and test the performance and reliability of the electron identification procedure, several samples in which these background processes are enhanced are studied.
\begin{itemize}
\item Electron misidentification in the central region is investigated by measuring the probability
of selecting a second electron in addition to the scattered electron in DIS candidate events when the
track quality criteria are relaxed. 
From a  sample of 244980 DIS candidates, 1563 events with a second electromagnetic central cluster are selected if no tracking conditions are applied to this cluster.
The fake electrons are predominantly photons from DIS or Compton events (figure~\ref{control}a).
This contribution is greatly reduced by requiring a geometrical track--cluster match 
(figure~\ref{control}b). The remaining sample consists of 250 events. The fake electron background
is described by the simulation at the 20\% level. 
\item 
Electron misidentification in the forward region affects only the ``3e'' selection. The dominant contribution to this background is the misidentification of a hadron as an electron. 
The fake electron background in this region is studied with the DIS event sample in an analysis similar to that described above. 
Fake electrons in the forward region are also searched for in an inelastic Compton event sample with one electron and one photon in the central region. 
The fake electron background in the forward region is described by the simulation at the 20\% level. 
\item Detailed studies of photon conversions are performed using a sample enriched with elastic 
Compton events, selected by requiring one central electron plus a second 
central electromagnetic cluster (photon candidate) and no significant additional energy in the calorimeters.
Distributions of the charged tracks associated with the photon candidate are shown
in figures~\ref{control}c~and~\ref{control}d. 
The number of tracks (figure ~\ref{control}c) and their starting radius  
(figure~\ref{control}d)
are well reproduced by the simulation.  
In figure~\ref{control}d, the central tracker structure is 
visible as peaks in the distribution, corresponding to photon conversions 
in the tracker walls. The first peak is populated by tracks associated with true electrons and by tracks from conversions which occur before the active tracker volume. The second peak is due to photon conversions in the dead material between the inner and outer central trackers. These conversions are described by the simulation to better than 20\%.  
\end{itemize}
Based on those studies, the uncertainty on the background simulation is estimated to be 20\%.

\subsection{Systematic uncertainties }

The systematic uncertainties are related to the measurement of the electron pair production process, to the theoretical description of this process and to the background simulation.
\par 
The main measurement uncertainty is due to the tracking conditions in the electron identification procedure. 
The electron track association efficiency 
 is measured with a DIS sample selected with $E-P_z>45$~GeV and 
a single electromagnetic cluster in the calorimeter with a transverse 
momentum above 10~GeV in the polar angle interval $20^\circ < \theta < 150^\circ$. 
The measured average track association efficiency is 90\% and varies only slightly with the track momentum and polar angle. 
This efficiency, measured to a precision
ranging from 3\% for polar angles around $90^\circ$ to 15\% at the forward edge of the
angular acceptance of the central tracker ($\theta=20^\circ$), is well described by the simulation. 
Uncertainties on the energy scales of the calorimeters, on the trigger 
efficiency and on the luminosity measurement are also taken into account.
The total measurement uncertainty is typically  7\% for the ``2e'' selection and 10\% for the ``3e'' selection. 
\par
The theoretical uncertainty on the pair production process cross section, calculated with GRAPE, is 3\%, as explained in  section~\ref{theory}.
The uncertainty on the Compton and DIS background contributions is 20\%, deduced from the studies presented in  section~\ref{bg}. 
\par
The error on the event yields predicted by GRAPE (section~\ref{yields}) contains all measurement and theoretical errors added in quadrature. The uncertainty on the total SM prediction also includes the errors on
the Compton and DIS backgrounds.
\par
The error on the extracted cross sections (section~\ref{xsec}) includes all measurement and background errors as described above. Theoretical errors are applied to the GRAPE prediction of the cross sections.

\section{Results}

\subsection{Multi--electron event samples}
\label{yields}
The multi--electron event yields are summarised in table~\ref{metab1}.
The observed numbers of ``2e'' and ``3e'' events are in agreement 
with the expectations, as is the number of events in the ``$\gamma\gamma$'' sample. No event is found with four or more identified electrons.

\begin{figure}[p] 
  \begin{center} \vspace*{-.5cm}
      \epsfig{file=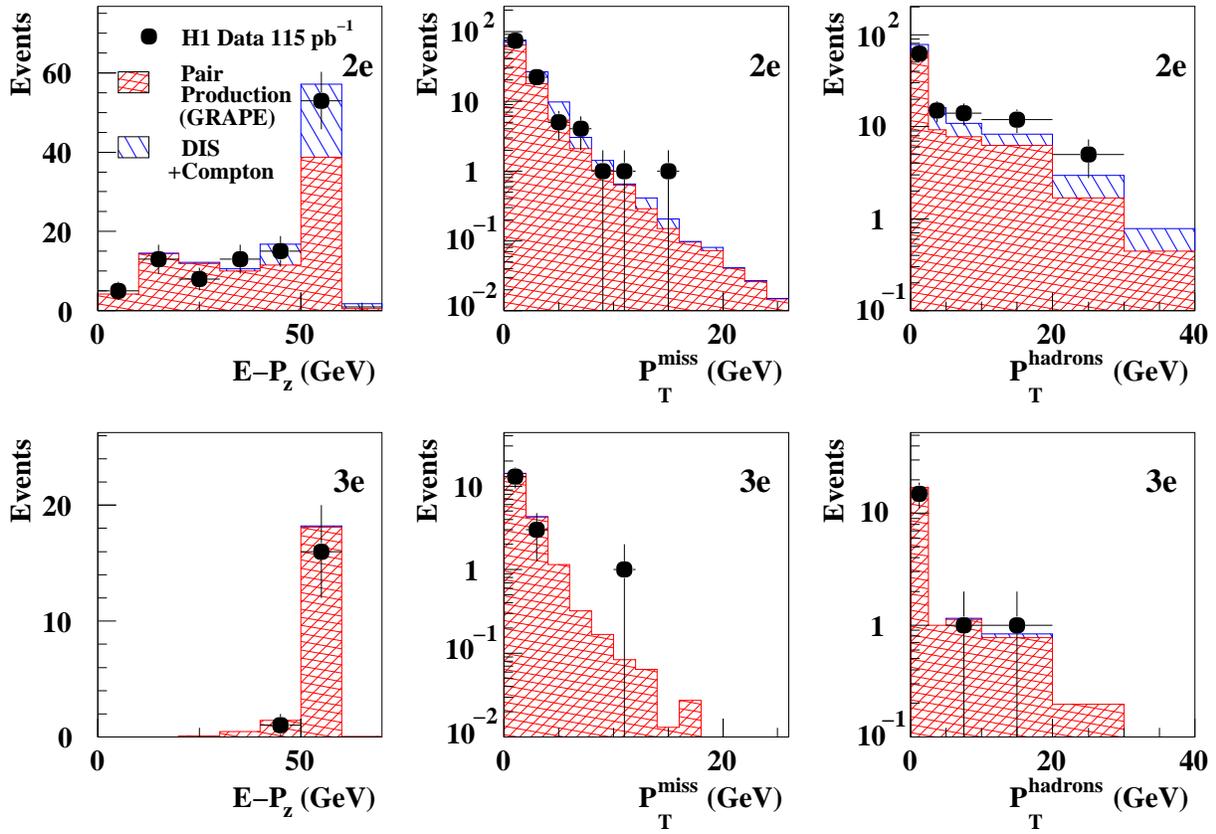,width=16cm}
  \end{center} \vspace*{-.3cm}
  \caption{  Distributions of $E-P_z$, $P_T^{miss}$ and $P_T^{hadrons}$ for events classified as
``2e'' (top) and ``3e'' (bottom), compared with  expectations.}
  \label{global_variables}
\end{figure}

The distributions of longitudinal momentum balance $E-P_z$, missing transverse momentum $P_T^{miss}$ and hadronic transverse momentum $P_T^{hadrons}$ 
are presented in figure~\ref{global_variables}. 
The ``3e'' events accumulate at $E - P_z$ values around 55 GeV, 
as expected if the scattered electron is visible in the detector.
The ``2e'' events show a tail at lower $E - P_z$, 
due to the scattered electron being lost in the beam pipe,
corresponding to the dominant $\gamma\gamma$ topology. 
The missing transverse momentum $P_T^{miss}$ is taken to be the modulus of the vector 
sum of the transverse momenta of all visible particles.
The  $P_T^{miss}$ distributions are consistent with the expectation for no
emission of undetected particles with substantial transverse momentum.
The spectrum of the transverse momentum $P_T^{hadrons}$ 
of all visible particles except identified electrons is also well 
described by the SM prediction.
\par
\begin{figure}[htb]
  \begin{center}
     \epsfig{file=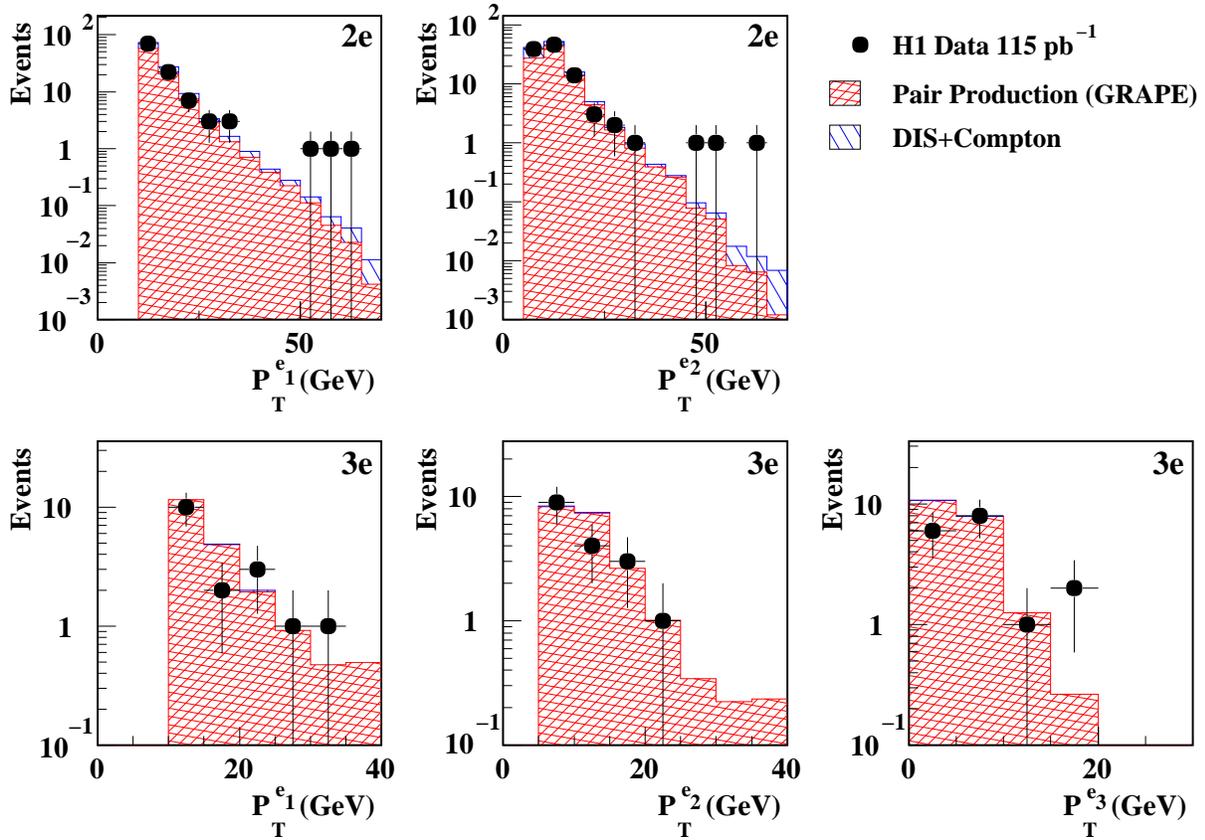,width=16cm}%
  \end{center}
  \caption{ Electron transverse momentum distributions 
for events classified as ``2e'' (top) and ``3e'' (bottom) compared with expectations. Electrons are ordered according to decreasing transverse momentum, $P_T^{e_i} > P_T^{e_{i+1}}$.}
  \label{transverse_momenta}
\end{figure}

The distributions of the individual electron transverse momenta $P_T^{e_i}$ 
are steeply falling as shown in figure~\ref{transverse_momenta}. 
The ``2e'' and ``3e'' samples are in good overall agreement with the SM, except for three ``2e'' events 
 with $P_T^{e_1}$ above 50~GeV, where the SM expectation 
is small. 
\par
\begin{figure}[htb]
  \begin{center}
     \epsfig{file=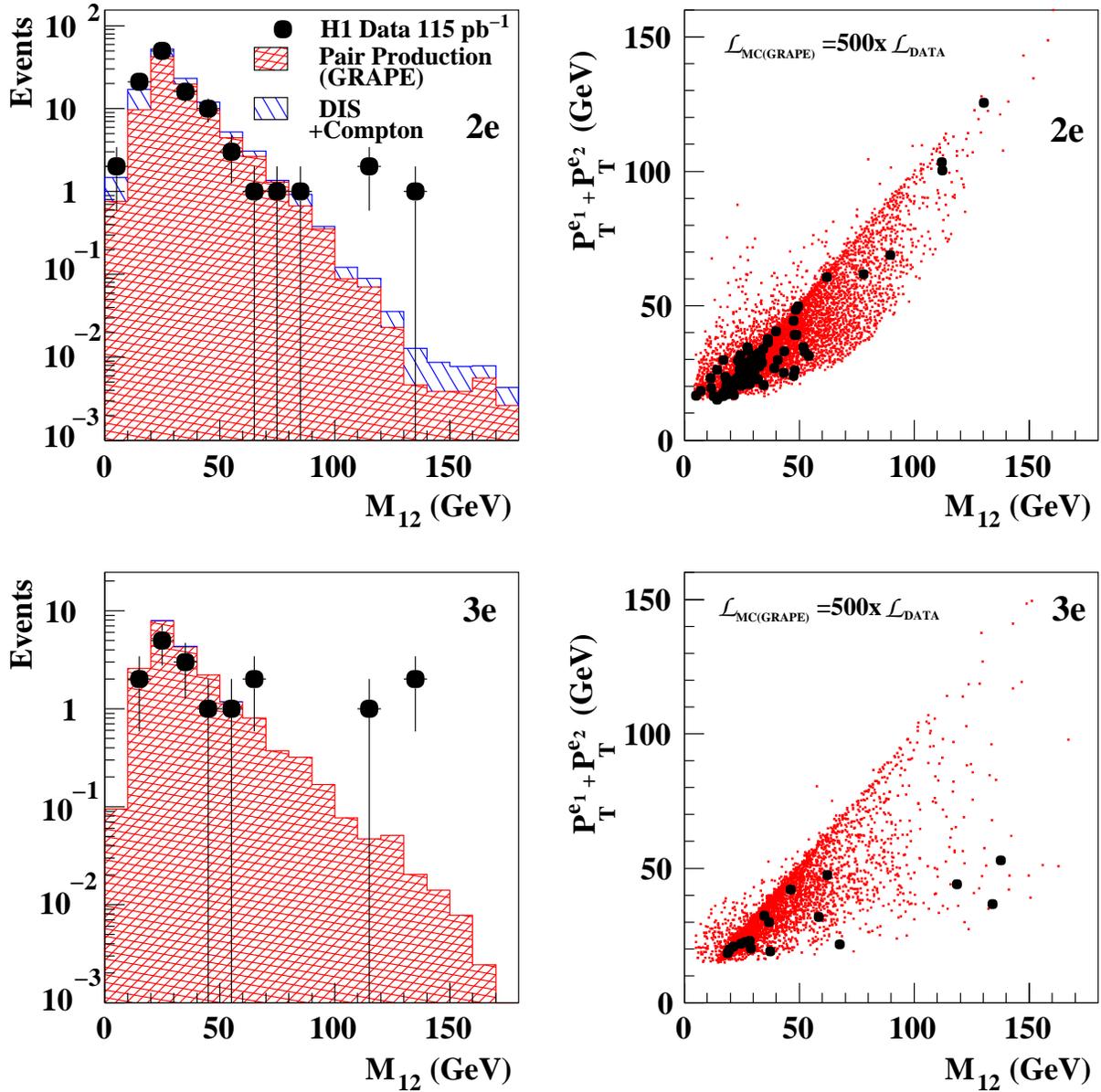,width=16cm}%
  \end{center}
  \caption{ Distribution of the invariant mass $M_{12}$ of the two 
highest $P_T$ electrons compared with expectations (left) and the correlation of $M_{12}$ with
the scalar sum of the $P_T$'s of the electrons (right) for events 
classified as ``2e'' (top) and ``3e'' (bottom). 
The bold dots on the right plots represent the data while the small points represent the pair production (GRAPE) prediction for a luminosity 500 times higher than that of the data.}
  \label{m12}
\end{figure}
\begin{figure}[htb]
  \begin{center}
     \epsfig{file=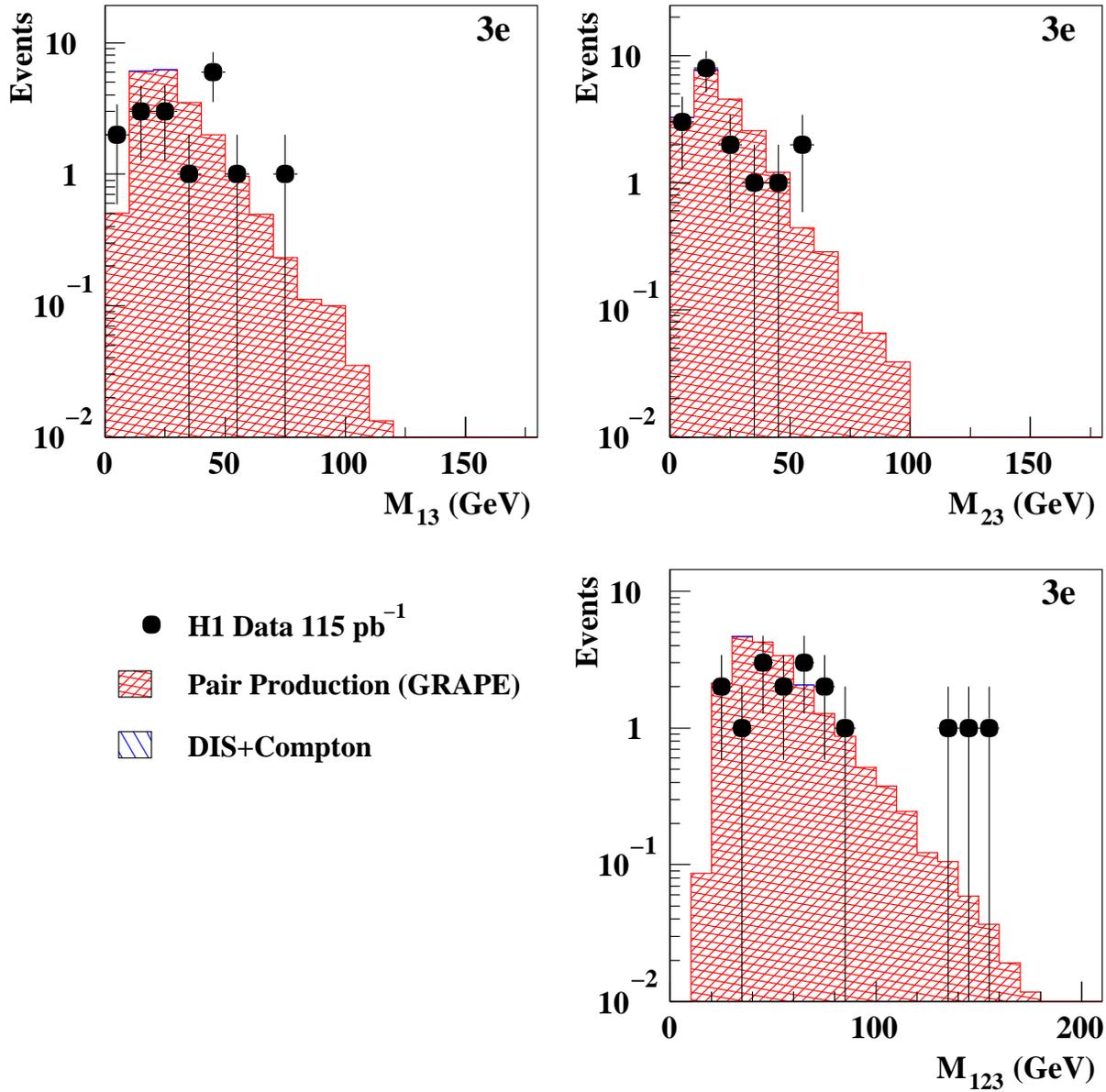,width=16cm}%
  \end{center}
  \caption{ Invariant mass distributions of electron pairs 1-3
and 2-3 (top left and right) and of the tri--electron system (bottom right)
for events classified as ``3e''.}
  \label{m123}
\end{figure}

The distribution of the invariant mass of the two highest 
$P_T$ electrons in the event ($M_{12}$) and the correlation with the scalar sum
$P_T^{e_1} + P_T^{e_2}$ are shown in figure~\ref{m12}. The agreement with the SM prediction is good at low $M_{12}$.
However, three ``2e'' and three ``3e'' events are seen with
invariant masses $M_{12}$ above 100~GeV, where the SM expectation is small.
The three ``2e'' events are the same as those observed at high $P_T^{e_1}$.
The invariant masses $M_{13}$, $M_{23}$ and $M_{123}$ of the other possible electron combinations 
in the ``3e'' sample are shown in figure~\ref{m123}. 
No event is seen with an unexpectedly high $M_{13}$ or $M_{23}$.
The three high $M_{12}$ events also give rise to the largest 
tri--electron masses $M_{123}$.
The comparison of the observed events
with masses  $M_{12}$ above 100~GeV  with the SM expectations is presented in table~\ref{metab2}. These events are discussed in detail in section~\ref{section:events}.

\subsection{Cross section measurement }
\label{xsec}

Using the selected ``$\gamma\gamma$'' sample, electron pair production cross sections are measured in the kinematic region 
defined by 
$$
20^\circ < \theta^{e_{1,2}} < 150^\circ, \;\;\;
P_T^{e_1}>10~\mathrm{GeV}, \;\;\; 
P_T^{e_2}>5~\mathrm{GeV}, \;\;\;
y<0.82, \;\;\;\;
Q^2<1~\mathrm{GeV}^{2}\;.
$$
For this measurement, the data samples collected at $\sqrt{s}$=301~GeV and 319~GeV are combined taking into account their respective luminosities. 
Assuming a linear dependence of the cross section on
 the proton beam energy, as predicted by the SM, the resulting cross section corresponds to an effective $\sqrt{s}$ = 313~GeV.
\par
After background subtraction, the observed number of events is corrected
for acceptance and detector effects to obtain the cross section.
The generator GRAPE is used to calculate the detector acceptance $A$
for this region of phase space. The acceptance accounts for 
detection efficiencies and migrations. The cross section is thus
$$
  \sigma = \frac{{\rm N_{data}} - {\rm N_{bgd}}}{{\cal L}{A}},
$$
where ${\rm N_{data}}$ is the number of events observed, ${\rm
  N_{bgd}}$ is the number of events expected from background processes (DIS and Compton) and
${\cal L}$ is the integrated luminosity of the data sample.
 
\begin{figure}[htb]
  \begin{center}
     \epsfig{file=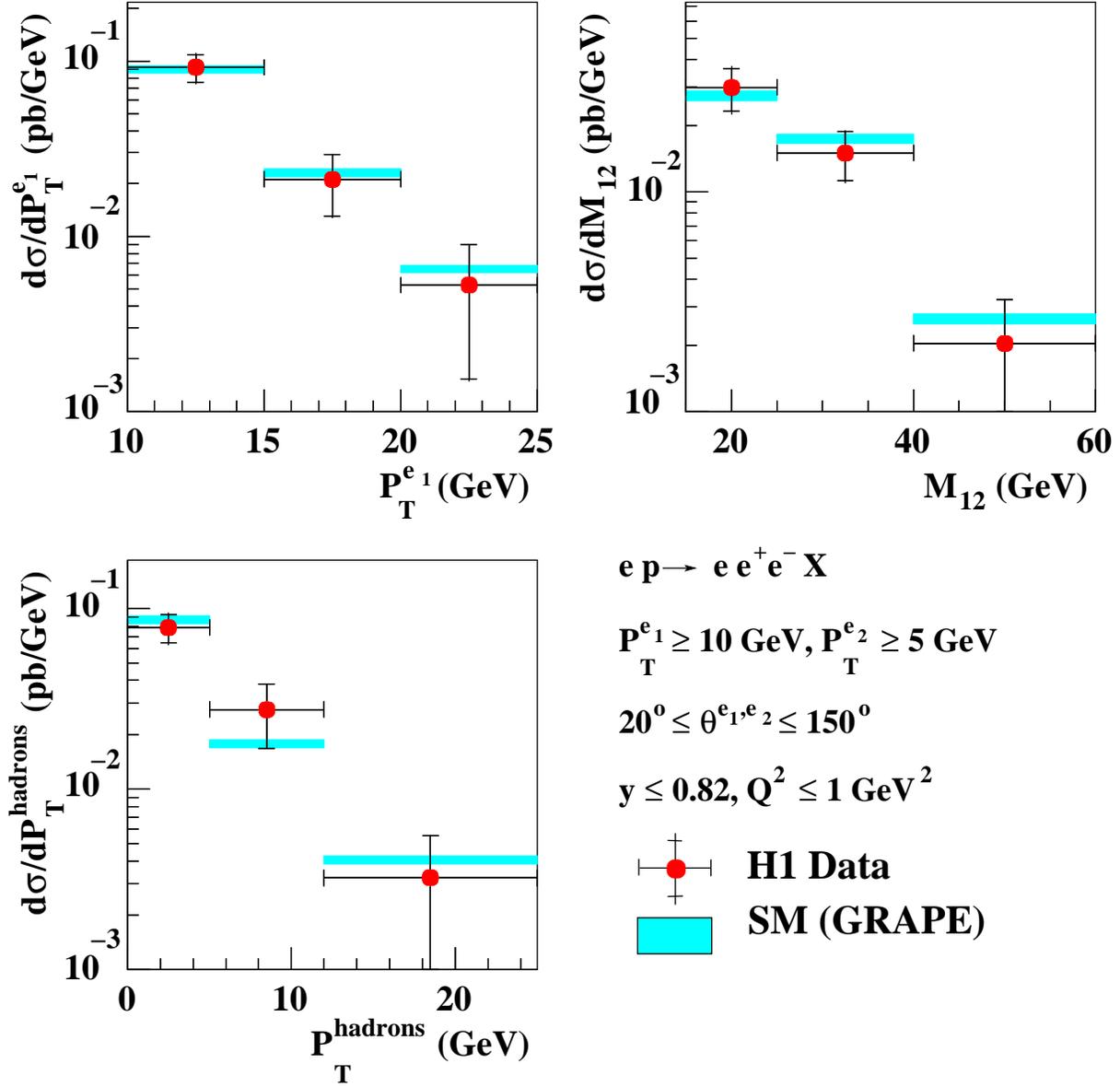,width=16cm}%
  \end{center}
  \caption{ Cross section measurements in a restricted phase space dominated by  the photon--photon process
as a function of the leading electron transverse momentum ($P_T^{e1}$), the invariant mass of the electron--positron pair ($M_{12}$) and  the hadronic transverse momentum ($P_T^{hadrons}$). The differential cross sections are averaged over the intervals shown. 
The inner error bars on the data points represent the statistical error, which dominates the measurement uncertainty. The outer error bars show the statistical and systematic uncertainties added in quadrature. The bands represent the one standard deviation uncertainty in the SM prediction. }
  \label{sigma_xsec}
\end{figure}

The $ep\rightarrow ee^+e^-X$ cross section, integrated over the phase space defined above, is
$$
  \sigma = ( 0.59 \pm 0.08 \pm 0.05 ) {\rm~pb},
$$
where the first error is statistical and the second  systematic, obtained as described in section 4.2.
This result agrees well with the SM  expectation of $( 0.62\pm 0.02 )$~pb, calculated with GRAPE. The differential cross sections as a function of $P_T^{e_1}$, $M_{12}$ and $P_T^{hadrons}$
 are shown in figure~\ref{sigma_xsec} and table~\ref{metab3}. The measurements are in good agreement with the expected cross sections.

\subsection{Discussion of high mass events}
\label{section:events}
\begin{figure}[htb]
  \begin{center}
     \epsfig{file=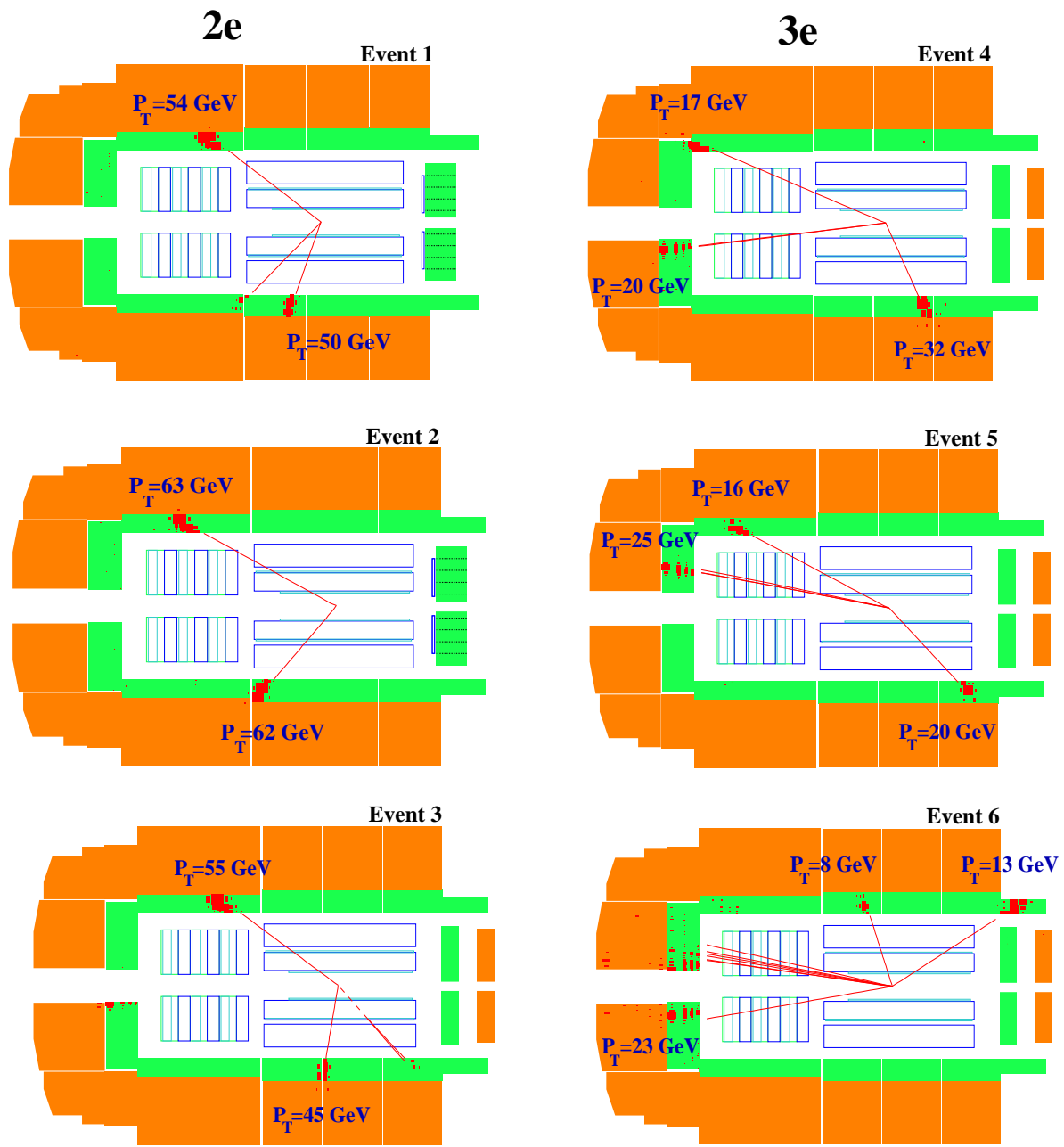,width=16cm}%
  \end{center}
  \caption{Displays of the three ``2e'' events (left) and the three~~``3e''~~events (right)~~with $M_{12}>100$~GeV
in the $R - z$ view. 
The reconstructed tracks and the energy deposits in the 
calorimeters are indicated (see text and table~\ref{tab_kin}).
The beam positrons enter the detector from the left and the protons from the right.}
  \label{meevdall}
\end{figure}

All six events with $M_{12}>100$~GeV were recorded during 
positron--proton collisions. 
For these events, displayed in figure~\ref{meevdall}, all available detector information supports
the interpretation of the electron candidates as being true electrons.
The electromagnetic shower shapes are checked individually 
and found to  be similar to those expected from the calorimeter response to electrons.  
All central tracks yield a specific ionisation
in the central drift chamber  as expected for  single electrons.
The measurements of the central electron momenta by the 
tracker and the calorimeter are compatible within errors.
The forward electron candidates in the ``3e'' events 4, 5 and 6 all have at
least one track pointing to the calorimetric energy cluster, although no such
requirement is made in the identification procedure.  
\par
Although classified as ``2e'', event 1 also contains a third
electron candidate with energy below the identification threshold.
Similarly, event 3 has a compact electromagnetic energy deposit 
located at the forward boundary of the liquid argon calorimeter, 
outside the electron identification fiducial volume. This event also
contains a low energy converted photon radiated close to electron $e_2$.
With the exception of event 6, which has a high energy forward hadronic jet, the events show no hadronic activity in the detector.
It should be noted that the 
high mass di--electron  topology differs for the observed events classified 
as ``2e'' and ``3e''. In the former case, the high--mass 
is formed from two central high--$P_T$ electrons,
whereas in the latter it is formed from one forward 
and one central electron, both of intermediate $P_T$ (figure~\ref{m12}). 
\par
The event kinematics of the six high mass events are presented in table~\ref{tab_kin}. The electron energy, electron polar angle and forward and backward electron azimuthal angles are measured from the calorimetric deposits. For the central electrons, the azimuthal angle is measured from the associated track, which yields a better precision. The electric charge of the electrons,  measured in the central region from the associated track curvature, is given in table~\ref{tab_kin} if the significance of its determination exceeds two standard deviations.  All events are compatible with the presence of one $e^-$ and two $e^+$ in the final state, as expected from pair production process.
\par
Imposing longitudinal and transverse momentum conservation, a constrained fit can be performed to improve the kinematic measurement. This corresponds to an adjustment of particle observables (energy, polar and azimuthal angles) within experimental errors in order to achieve $E-P_z=55.2$~GeV and $P_T^{miss}=0$~GeV. In event 2 only two electrons are visible  and the  measured $E-P_z$ value is significantly lower than 55 GeV. For this event it is supposed that the scattered electron has escaped down the beampipe and therefore the $E-P_z$ constraint is removed in the kinematic fit.  The $M_{12}$ values obtained from  the kinematic fit are indicated in table~\ref{tab_kin}. They are consistent with the direct measurements. The errors are reduced by more than a factor of two with the exception of event 2. The  $\chi^2$ per degree of freedom of the kinematic fit is in the range 0.7 to 1.7 for the six events, showing that the kinematics of the six high mass events are well understood within the measurement errors.
The  $M_{12}$ values are incompatible with the interpretation of the six high mass electron pairs as the decay of a single narrow resonance.
The same is true for the $M_{123}$ values in the ``3e'' high mass events.

\section{Summary}

High--$P_T$ multi--electron production is measured for
the first time in $ep$ scattering at HERA. 
The di--electron and tri--electron event yields are in good overall agreement 
with the SM predictions. No events are observed with more than three identified electrons, again in agreement with the SM expectation.
 Differential cross sections  for electron pair production are extracted in a restricted phase space region  dominated by photon--photon 
interactions and
 are  found to agree with the predictions.

Within the di-- and tri--electron samples, the invariant mass of the two electrons with the highest transverse momenta is studied.
For masses above 100~GeV, three events classified as di--electrons and three events 
classified as tri--electrons are observed, 
compared to SM expectations of $0.30 \pm 0.04$ and
$0.23 \pm 0.04$, respectively. 

\section*{Acknowledgements}

We are grateful to the HERA machine group whose outstanding
efforts have made and continue to make this experiment possible. 
We thank
the engineers and technicians for their work in constructing and 
maintaining the H1 detector, our funding agencies for 
financial support, the
DESY technical staff for continual assistance, 
and the DESY directorate for support and the
hospitality which they extend to the non--DESY 
members of the collaboration. The authors wish to thank J.~A.~M.~Vermaseren and T.~Abe for many useful discussions.


\clearpage

\begin{table}[htb]
\begin{center}

\begin{tabular}{|l|c|c||c|c|c|}
\hline

   Selection   & Data  &  SM &  Pair Production (GRAPE) &  DIS + Compton \\ \hline \hline

  ``2e''              &   108 &  $117.1\pm   8.6$ &  $ 91.4\pm   6.9$ &  $ 25.7\pm   5.2$ \\
  ``3e''              &    17 &  $ 20.3\pm   2.1$ &  $ 20.2\pm   2.1$ &  $  0.1\pm   0.1$ \\
 ``4e'' or more               &     $0$ &  $ 0.12\pm  0.04$ &  $ 0.12\pm  0.04$ & $ <0.02$ {\footnotesize (95\% C.L.)}  \\
   \hline
 "$\gamma\gamma$" subsample      &    42 &  $ 44.9\pm   4.2$ &  $ 43.7\pm   4.2$ &  $  1.2\pm   0.4$ \\
   \hline
  
 \end{tabular}

\caption{
Observed and predicted multi--electron event yields  for the samples described 
in the text. The analysed data sample corresponds to an integrated luminosity of $115.2$~pb$^{-1}$.
The  errors on the predictions include model uncertainties and experimental systematic errors added in quadrature.
}
\label{metab1}
\end{center}
\end{table}

\begin{table}[htb]
\begin{center}
 
\begin{tabular}{|c|c|c||c|c|c|}
\hline
   Selection   & Data  &  SM &  Pair Production (GRAPE) &  DIS + Compton \\ \hline \hline

 ``2e''   $M_{12}>100$~GeV &     $3$ &  $ 0.30\pm  0.04$ &  $ 0.21\pm  0.03$ &  $ 0.09\pm  0.02$ \\
 ``3e''   $M_{12}>100$~GeV &     $3$ &  $ 0.23\pm  0.04$ &  $ 0.23\pm  0.03$ &  $ <0.02$ {\footnotesize ( 95\% C.L.)} \\
   \hline
 \end{tabular}
 
\caption{ Observed and predicted multi--electron event yields for 
masses $M_{12}>100$~GeV for the samples described in the text.
The analysed data sample corresponds to an integrated luminosity of $115.2$~pb$^{-1}$.
The errors on the predictions include model uncertainties and experimental 
systematic errors added in quadrature.}
\label{metab2}
\end{center}
\end{table}

\begin{table}[htb]
\begin{center}
 \renewcommand{\arraystretch}{1.23}  
\begin{tabular}{|c|c|c|}
\hline
   Variable range   & Measured cross section  &  Pair Production (GRAPE) cross section  \\ 
   $[$GeV$]$   & $[$pb/GeV$]$  &  $[$pb/GeV$]$ \\ \hline \hline
\multicolumn{3}{c}{$d\sigma/dP_T^{e_1}$} \\ \hline
$   10< P_T^{e_1}      <   15$   &   $  0.092\pm  0.016\pm  0.009$   &    $ 0.090\pm  0.003$    \\\hline
$   15< P_T^{e_1}      <   20$   &   $  0.021\pm  0.008\pm  0.002$   &    $ 0.023\pm  0.001$    \\\hline
$   20< P_T^{e_1}      <   25$   &   $ 0.0053\pm 0.0037\pm 0.0007$   &    $0.0065\pm 0.0002$    \\\hline
\hline
\multicolumn{3}{c}{$d\sigma/dM_{12}$} \\ \hline
$   15< M_{12}         <   25$   &   $  0.030\pm  0.007\pm  0.003$   &    $ 0.027\pm  0.001$    \\ \hline
$   25< M_{12}         <   40$   &   $  0.015\pm  0.004\pm  0.001$   &    $ 0.017\pm  0.001$    \\ \hline
$   40< M_{12}         <   60$   &   $ 0.0020\pm 0.0012\pm 0.0002$   &    $0.0026\pm 0.0001$    \\ \hline
\hline
\multicolumn{3}{c}{$d\sigma/dP_T^{hadrons}$} \\ \hline
$    0< P_T^{hadrons}  <    5$   &   $  0.079\pm  0.014\pm  0.009$   &    $ 0.087\pm  0.003$    \\ \hline
$    5< P_T^{hadrons}  <   12$   &   $  0.028\pm  0.011\pm  0.002$   &    $ 0.018\pm  0.001$    \\ \hline
$   12< P_T^{hadrons}  <   25$   &   $ 0.0032\pm 0.0023\pm 0.0005$   &    $0.0041\pm 0.0001$    \\ \hline

   \hline
 \end{tabular}
 
\caption{ Differential cross sections 
$d\sigma/dP_T^{e_1}$, $d\sigma/dM_{12}$ and $d\sigma/dP_T^{hadrons}$ for the process $ep\rightarrow ee^+e^-X$ in a restricted phase space (see text). 
The differential cross sections are averaged over the quoted intervals. 
The first error is statistical and the second is systematic. Theoretical predictions with GRAPE are also shown. }
\label{metab3}
\end{center}
\end{table}

\begin{table}[htb]
\begin{center}
\footnotesize
\renewcommand{\arraystretch}{1.23}
\begin{tabular}{|c||c|c|c|c|}
\hline
Particle & E [GeV] & $\theta$ [degrees] & $\phi$ [degrees] & Charge (significance)\\
\hline \hline
\multicolumn{5}{|c|}{\bf Multi--electron Event 1 \;\;\;\;(2e)  } \\
\multicolumn{1}{|l}{Run  83507} & \multicolumn{2}{c}{ $E-P_z$ = 54.0 $\pm$ 1.1 GeV }& \multicolumn{2}{c|} {$P_T^{miss}$=3.1 $\pm$ 1.8 GeV}\\
\multicolumn{1}{|l}{Event  16817} & \multicolumn{2}{c}{$M_{12}$ = 111.2 $\pm$ 2.4 GeV}& \multicolumn{2}{c|} { $M_{12}^{fit}$ = 111.3 $\pm$ 0.4 GeV } \\
\hline
e$_1$ &  90.3 $\pm$ 3.1 & 36.6 $\pm$ 0.2 & 98.48 $\pm$ 0.05   & - (4$\sigma$) \\
e$_2$ &  53.6  $\pm$ 1.4 & 69.6 $\pm$ 0.3 & -77.05 $\pm$ 0.05  & undetermined \\
low energy e &  4.4  $\pm$ 0.3 & 44.3 $\pm$ 0.3 & -155.46 $\pm$ 0.03  & + (70$\sigma$) \\
\hline \hline
\multicolumn{5}{|c|}{ \bf  Multi--electron Event 2 \;\;\;\;(2e) } \\
\multicolumn{1}{|l}{ Run  89256} &\multicolumn{2}{c}{ $E-P_z$ = 43.9 $\pm$ 0.8 GeV}& \multicolumn{2}{c|} { $P_T^{miss}$=1.9 $\pm$ 1.8 GeV}\\
\multicolumn{1}{|l}{Event  224212} & \multicolumn{2}{c}{$M_{12}$ = 130.0 $\pm$ 2.6 GeV}& \multicolumn{2}{c|} { $M_{12}^{fit}$ = 129.3 $\pm$ 2.4 GeV } \\
\hline
e$_1$ &  132.4 $\pm$ 4.3 & 28.6 $\pm$ 0.1 & 8.73 $\pm$ 0.06  & undetermined \\
e$_2$ &  82.4  $\pm$ 1.8 & 48.4 $\pm$ 0.2 & -171.50 $\pm$ 0.03  & - (6$\sigma$) \\
\hline \hline
\multicolumn{5}{|c|}{ \bf  Multi--electron Event 3 \;\;\;\; (2e) } \\
\multicolumn{1}{|l}{ Run  254959} & \multicolumn{2}{c}{ $E-P_z$ = 57.3 $\pm$ 1.4 GeV}& \multicolumn{2}{c|} { $P_T^{miss}$=3.5 $\pm$ 2.0 GeV}\\
\multicolumn{1}{|l} {Event  17892} & \multicolumn{2}{c}{$M_{12}$ = 112.5 $\pm$ 2.4 GeV}& \multicolumn{2}{c|} { $M_{12}^{fit}$ = 109.5 $\pm$ 1.0 GeV } \\
\hline
e$_1$ &  96.9  $\pm$ 3.3 & 34.6 $\pm$ 0.3 & 52.66 $\pm$ 0.02  & + (10$\sigma$) \\
e$_2$ &  46.1   $\pm$ 1.1 & 80.1 $\pm$ 0.9 & -125.62 $\pm$ 0.01  & + (15$\sigma$) \\
fwd em cluster &  70$^{+100{  \;a}}_{-2}$ & 4.5 $\pm$ 0.1 & -132.7 $\pm$ 1.0  & undetermined\\
photon &  1.1   $\pm$ 0.1 & 132.0 $\pm$ 5.5 & 39.8 $\pm$ 7.3  & 0 \\ \hline
\multicolumn{1}{|l} {} &\multicolumn{4}{l|}{\footnotesize $^a$ this error includes the uncertainty due to energy loss in the beampipe}\\
\hline \hline
\multicolumn{5}{|c|}{ \bf  Multi--electron Event 4 \;\;\;\; (3e)} \\
\multicolumn{1}{|l} {Run  168058} &\multicolumn{2}{c}{ $E-P_z$ = 55.7 $\pm$ 1.4 GeV}& \multicolumn{2}{c|} { $P_T^{miss}$=1.1 $\pm$ 0.8 GeV}\\
\multicolumn{1}{|l}{Event  42123} & \multicolumn{2}{c}{$M_{12}$ = 137.4 $\pm$ 2.9 GeV}& \multicolumn{2}{c|} { $M_{12}^{fit}$ = 138.8 $\pm$ 1.2 GeV} \\
\hline
e$_1$ &  35.8  $\pm$ 0.9 & 115.6 $\pm$ 0.9 & -5.98 $\pm$ 0.02  & + (18$\sigma$) \\
e$_2$ &  173.0 $\pm$ 5.5 & 6.6 $\pm$ 0.1 & -159.1 $\pm$ 0.5  & undetermined \\
e$_3$ &  44.8  $\pm$ 1.7 & 21.8 $\pm$ 0.2 & 139.10 $\pm$ 0.03  & - (12$\sigma$) \\
\hline \hline
\multicolumn{5}{|c|}{ \bf  Multi--electron Event 5 \;\;\;\; (3e)} \\
\multicolumn{1}{|l}{ Run  192864} &\multicolumn{2}{c}{ $E-P_z$ = 53.8 $\pm$ 1.4 GeV}& \multicolumn{2}{c|} { $P_T^{miss}$=0.7 $\pm$ 0.6 GeV}\\
\multicolumn{1}{|l}{ Event  123614} & \multicolumn{2}{c}{ $M_{12}$ = 118.1 $\pm$ 2.6 GeV}& \multicolumn{2}{c|} { $M_{12}^{fit}$ = 121.9 $\pm$ 0.6 GeV} \\
\hline
e$_1$ &  138.9 $\pm$ 4.5 & 10.2 $\pm$ 0.1 & 44.1 $\pm$ 0.6  & undetermined \\
e$_2$ &  28.1   $\pm$ 0.8 & 134.7 $\pm$ 0.3 & -95.85 $\pm$ 0.01  & + (25$\sigma$) \\
e$_3$ &  35.3  $\pm$ 1.5 & 26.6 $\pm$ 0.1 & 172.71 $\pm$ 0.05  & + (5$\sigma$) \\
\hline \hline
\multicolumn{5}{|c|}{\bf  Multi--electron Event 6 \;\;\;\; (3e)} \\
\multicolumn{1}{|l}{ Run  267312 } & \multicolumn{2}{c}{ $E-P_z$ = 57.4 $\pm$ 1.6 GeV}& \multicolumn{2}{c|} { $P_T^{miss}$=2.4 $\pm$ 0.8 GeV}\\
\multicolumn{1}{|l}{Event  203075} & \multicolumn{2}{c}{$M_{12}$ = 134.7 $\pm$ 3.1 GeV}& \multicolumn{2}{c|} { $M_{12}^{fit}$ = 132.3 $\pm$ 1.4 GeV} \\
\hline
e$_1$ &  186.0  $\pm$ 5.9 & 7.11 $\pm$ 0.05 & -71.3 $\pm$ 0.4  & undetermined \\
e$_2$ &  25.5   $\pm$ 0.8 & 148.8 $\pm$ 0.2 & 120.25 $\pm$ 0.02  & + (32$\sigma$) \\
e$_3$ &  8.5   $\pm$ 0.5 & 69.7 $\pm$ 0.3 & 164.90 $\pm$ 0.01  & + (57$\sigma$) \\
hadrons$^b$ &  123.2   $\pm$ 6.7 & 6.1 $\pm$ 1.1 & 53.5 $\pm$ 1.1  &  \\ \hline
\multicolumn{1}{|l} { } & \multicolumn{4}{l|}{\footnotesize $^b$ mass of the visible hadronic system: 24.0 $\pm$ 2.5 GeV}\\
\hline 
\end{tabular}
\caption[]
  {\label{tab_kin} Reconstructed kinematics of the six multi-electron events with $M_{12}>100$~GeV (see text).
$E$ is the particle's energy and $\theta$ and $\phi$ its polar and azimuthal angles, respectively .
The charges of the electron candidates are also given, if they are measured with a significance of better than two standard deviations. The six events were recorded in $e^+p$ collisions.\\

}
\end{center}
\end{table}

\newpage
%

\end{document}